\documentstyle[12pt]{article}

\begin{document}

\newcommand{\sls}[1]{#1 \! \! \! /}
\newcommand{\SLS}[1]{#1 \! \! \! \! /}
\newcommand{\open}{\langle \langle}
\newcommand{\close}{\rangle \rangle}
\newcommand{\fatpi}{ {\bf \Pi} }
\newcommand{\der}{\partial}

\begin{titlepage}
\begin{center}
\today  \hfill  VPI-IPPAP-96-5 \\
\vskip .5 in

{\LARGE The Bilocal Effective Action for}
\vskip .1 in
{\LARGE Mesons in Large-N QCD}

\vskip .3in

{\large Gregory L. Keaton}
\vskip .45in

{\it Institute for Particle Physics and Astrophysics \\}
{\it Physics Department, Viginia Tech, Blacksburg VA  24061-0435 \\}
{\it keaton@mustang.phys.vt.edu}
\end{center}

\vskip .5in

\begin{abstract}
The bilocal effective action is developed as a viable tool to study
mesons in the large-N limit of QCD.  Several results from current
algebra are derived in a new way using this action.  The question
of gauge invariance is discussed in an appendix
\end{abstract}

\end{titlepage}

%\newpage
%\renewcommand{\thepage}{\arabic{page}}
%\setcounter{page}{1}

\section{Introduction}

Although quantum chromodynamics has been extremely successful at 
describing quarks and gluons perturbatively,
it has yet to make many firm predictions about
mesons and baryons.  How can the 
composite states be described, starting from the
Lagrangian for the constituents?  Many attempts invoke the
constituent quark model \cite{cq} and/or the potential or 
bag model \cite{dgh}, none
of which
can be derived from QCD.  
The Nambu--Jona-Lasinio model offers another approach \cite{njl}, at the
expense of approximating the gluon propagator by a delta-function.

Perhaps the most puzzling feature of these models is that they work so well.
Surely these models' successes, in spite of gross dynamical approximations,
indicate that each model shares with QCD a common symmetry
(or other fundamental principle) that is actually dictating the favorable
results.  Those predictions that depend explicitly on the dynamics
are (perhaps) the less successful ones.
However, it is impossible to tell which of the models' predictions are
artifacts of the dynamical assumptions, and which result from the fundamental
principles, since the dynamical approximations are employed from the
beginning.

It therefore seems worthwhile to attempt a systematic approach to
deriving hadron properties from QCD, making as few assumptions as possible.
To do this, I have concentrated on the mesons, and have used the
effective action for bilocal operators. 

However, some approximation must still be used.  I have used the large number of
colors (large-N) expansion, which offers a formal justification for the
``quenched'' approximation.  In this scheme, a meson consists of a single
quark--anti-quark pair (and an indefinite number of gluons), other virtual 
quark pairs being suppressed by powers of $1/N$ \cite{1/N}.  
Therefore the meson n-point functions contain a single fermion
loop; this simplification allows one to write down---at least formally---
exact expressions for these functions.  And even though the n-point functions
still cannot be computed, it is possible to derive relations between them.

In this paper, the effective action for large-N QCD is derived.
The vacuum is discussed first, then the mesons.  The approach
illuminates the dual nature
of the pion:  it is both a bound state solution to the
Bethe-Salpeter equation, and it is a Goldstone boson, present because
the vacuum state spontaneously breaks chiral symmetry.  Finally, heavier
mesons are considered.  A little-known sum rule is derived in a new
way using this approach; the sum rule agrees quite well with experiment.
The appendix addresses the question of the gauge
invariance of the results.

\section{The Effective Action}

We begin with the formalism of Cornwall, Jackiw, and Tomboulis \cite{cjt}.
They derive an effective action for the bilocal operator
$G_{ab}(x,y) = \langle T \psi_a (x) \bar{\psi}_b (y) \rangle$.  The
indices $a$ and $b$ give the spin, isospin, and color of the fermion;
they are suppressed below.  The effective action is
\begin{equation}
\Gamma[G] = i \, \mbox{tr log}[G] - \mbox{tr}[(i \sls{\partial}  - m)G]
 + \Gamma_2[G]
\label{jeffa}
\end{equation}
The matrix $m$ gives the current quark masses ($\sim$ 5 to 10 MeV for the $u$
and $d$ quarks).  $\Gamma_2[G]$ is the sum of the two-particle irreducible
(2PI)
diagrams, such as in Fig.\ 1.  That is, it is the
sum over all diagrams that remain connected after 
breaking any two fermion lines.  The fermion propagator for these
graphs is equal to $G$.
$\Gamma_2$ can be expressed formally as
\[
i\Gamma_2 = \frac{i^2}{2!} \langle \langle T (\bar{\psi} \sls{A} \psi)
 (\bar{\psi} \sls{A} \psi) \rangle \rangle - \frac{i^3}{3!}
 \langle \langle T (\bar{\psi} \sls{A} \psi) (\bar{\psi} \sls{A} \psi)
 (\bar{\psi} \sls{A} \psi)  \rangle \rangle + \ldots
\]
where the double brackets mean that the path integral over all possible
field configurations must be performed, and in the end 
only the 2PI diagrams are retained.  The sum over possible $A_{\mu}$'s
is subject to a gauge-fixing constraint; the appendix shows that
physical observables are, however, gauge invariant.

To leading order in the $1/N$ approximation, there is only one fermion loop
in each diagram \cite{1/N}; therefore, replacing all fermion propagators by
$G$, including the appropriate symmetry factors, and remembering an overall
minus sign for the fermion loop,
\begin{eqnarray*}
i \Gamma_2[G]  & = & - \left[ \frac{i^2}{2} \mbox{tr} \langle \langle G \sls{A}
 G \sls{A}
 \rangle \rangle - \frac{i^3}{3}  \mbox{tr} \langle \langle
 G \sls{A} G \sls{A}  G \sls{A} \rangle \rangle + \ldots \right] \\
 & = & \langle \langle \mbox{tr log} (1 + i G \sls{A}) \rangle \rangle \\
 & = & \langle \langle \mbox{tr log} \, i \left( G - \frac{i}{\sls{A}}  \right)
 \rangle \rangle + \langle \langle \mbox{tr log} \sls{A} \rangle \rangle
\end{eqnarray*}
Therefore, ignoring any constants independent of $G$, the total action is
\begin{equation}
\Gamma[G] = i \, \mbox{tr log} [G] - \mbox{tr}[(i \sls{\partial} - m) G]
 - i \open \mbox{tr log} \left( G - \frac{i}{\sls{A}} \right) \close
\label{neffa}
\end{equation}

\section{The Ground State}

The function $G$ satisfies the equation of motion $\delta \Gamma / \delta G
= 0$:
\begin{equation}
i G^{-1} - (i \sls{\partial} - m) - i \open \frac{1}{G - \frac{i}{\sls{A}}}
 \close = 0
\label{three}
\end{equation}
The rest of this section does no more than define some new notation
so that Eq.(\ref{three}) can be written more conveniently as
Eq. (\ref{four}), below.

In the ground state, translation invariance requires that $G(x,y)$ be a function
of $x-y$ only:  $G(x,y) = G_0(x-y)$.  Using parity conservation and Lorentz
invariance, $G_0$ can be expressed in terms of $p$, the momentum conjugate
to $x-y$, as
\[
i G_0^{-1} = a(p^2) \sls{p} - b(p^2)
\]
for two unknown functions $a$ and $b$.  If $b \neq 0$ when the quark mass
$m=0$, the propagator $G_0$ spontaneously breaks chiral symmetry.  Coleman
and Witten have shown \cite{cw} that large-N QCD does in fact break chiral
symmetry spontaneously, so we expect that $b \neq 0$.

Defining $\alpha \equiv a/(a^2p^2 - b^2)$ and $\beta \equiv b/(a^2p^2 - b^2)$,
we can write
\[
G_0 = i(\alpha \sls{p} + \beta) .
\]
Then Eq. (\ref{three}) becomes
\begin{equation}
(a-1) \sls{p} - (b-m) = \open \frac{1}{\alpha \sls{p} - \frac{1}{\sls{A}}
+ \beta } \close
\label{star}
\end{equation}
The notation can be further condensed by defining $\SLS{P} \equiv \alpha
\sls{p} - 1/\sls{A}$.  Then using the identity
\begin{equation}
\frac{1}{\SLS{P} \pm \beta} = \frac{1}{\SLS{P} \SLS{P} - \SLS{P} \beta
 \SLS{P}^{-1} \beta } ( \SLS{P} \mp \SLS{P} \beta \SLS{P}^{-1})
\label{dagger}
\end{equation}
and equating an even number of gamma-matrices on both the left- and right-hand
sides of Eq. (\ref{star}), we get
\begin{equation}
-b + m = - \open \frac{1}{\SLS{P} \SLS{P} - \SLS{P} \beta \SLS{P}^{-1} \beta}
 \SLS{P} \beta \SLS{P}^{-1} \close
\label{dagger2} 
\end{equation}
This expression is simplified by using Eq. (\ref{dagger}) to write
\begin{eqnarray*}
\frac{1}{\SLS{P} \SLS{P} - \SLS{P} \beta \SLS{P}^{-1} \beta}
 \SLS{P} \beta \SLS{P}^{-1} & = & \frac{1}{2} \left( \frac{1}{\SLS{P} - \beta}
 - \frac{1}{\SLS{P} + \beta} \right) \\
 & = & \frac{1}{\SLS{P}+\beta} \beta \frac{1}{\SLS{P}-\beta}
   =   \frac{1}{\SLS{P}-\beta} \beta \frac{1}{\SLS{P}+\beta} .
\end{eqnarray*}
Then Eq. (\ref{dagger2}) becomes
\begin{equation}
b = \open \frac{1}{\SLS{P}+\beta} \beta \frac{1}{\SLS{P}-\beta} \close + m.
\label{four}
\end{equation}
Because this equation gives a consistency condition for the order parameter
$b$ that appears as a result of the spontaneous breakdown of a symmetry, it
is analogous to the gap equation in the BCS theory of 
superconductivity.

\section{Excited States}

The previous section concerned the vacuum state where $G(x,y) = G_0(x-y)$.
Let us now look for the excited states of the theory.  They can be found by
allowing $G(x,y) = G_0(x-y) + \phi(x,y)$ and requiring that the total $G$
still satisfy the equation of motion, $\delta \Gamma / \delta G = 0$.
To see that this procedure is correct, note that it implies that
\[
0 = \left. \frac{\delta \Gamma}{\delta G} \right|_{G_0 + \phi}
 \approx  \: \left. \frac{\delta \Gamma}{\delta G} \right|_{G_0} +
 \phi \left. \frac{\delta^2 \Gamma}{\delta G^2} \right|_{G_0}
\]
The first term on the right hand side is zero by definition; using
Eq.(\ref{jeffa})
the second term gives
\begin{equation}
-i G_0^{-1} \phi G_0^{-1} + \phi \left. \frac{\delta^2 \Gamma_2}{\delta G^2}
\right|_{G_0} = 0.
\label{five}
\end{equation}
Since $\Gamma_2$ is the sum of 2PI graphs, $\delta^2 \Gamma_2/\delta G^2$ is
the irreducible four-point function.  Therefore Eq. (\ref{five}) is
simply the Bethe-Salpeter equation for $\phi$.  This confirms the
suspicion that
$\phi$ represents an excited state:  it is the Bethe-Salpeter
wavefunction \cite{bs} of the quark--anti-quark bound state.

We can now write the effective action for the meson fields $\phi$ by
expanding $\Gamma[G]$ around $G_0$:
\[
\Gamma[\phi]  =  
 - \frac{i}{2} \mbox{tr} \left( \phi \frac{1}{G_0} \phi
 \frac{1}{G_0} \right) + \frac{i}{3} \mbox{tr} \left( \phi \frac{1}{G_0}
 \phi \frac{1}{G_0} \phi \frac{1}{G_0} \right) - \ldots \;\;\;\;\;\;\;
\]
\begin{equation}
 + \frac{i}{2} \mbox{tr} \open \phi 
 \frac{1}{G_0 + \frac{i}{\sls{A}}}
 \phi \frac{1}{G_0 + \frac{i}{\sls{A}}} \close
 - \frac{i}{3} \mbox{tr} \open \phi \frac{1}{G_0 + \frac{i}{\sls{A}}}
 \phi \frac{1}{G_0 + \frac{i}{\sls{A}}} \phi \frac{1}{G_0 + \frac{i}{\sls{A}}}
 \close + \ldots 
\label{mesona}
\end{equation}
The equation of motion for $\phi$ is
\begin{equation}
\frac{1}{G_0} \phi \frac{1}{G_0} = \open \frac{1}{G_0 + \frac{i}{\sls{A}}}
\phi \frac{1}{G_0 + \frac{i}{\sls{A}}} \close + \mbox{(interactions)}
\label{six}
\end{equation}

\section{Pions}

Armed with this general formalism for mesons, we can look for particular
solutions to Eq. (\ref{six}).  It is easiest to start with pions, since we
expect them to exist on general grounds due to the spontaneous breakdown of
chiral symmetry.  With some foresight, then, we make the ansatz
\begin{equation}
\phi = \chi(x-y) \Pi^a(y) \gamma_5 T^a
\label{ansatz}
\end{equation}
for the pion field\footnote{As Bethe and Salpeter point out \cite{bs}, 
we could use $\phi = \chi(x-y) \Pi^a( \alpha x + (1-\alpha)y) \gamma_5 T^a$ 
for any $\alpha$ between 0 and 1.  The value $\alpha = 0$ is
chosen simply for convenience later.}.  ($T^a$ is a generator of the flavor
group.)  We expect that $\Pi^a(y)
\propto \exp(iky)$, where $k_{\mu}$ is the four-momentum of the pion
\cite{bs}.  For simplicity, consider the case when $k_{\mu} \rightarrow 0$
(which is only possible for physical pions if the mass of the pion is
zero).  Then Eq. (\ref{six}) becomes
\[
(a \sls{p} - b) \chi \gamma_5 T^a (a \sls{p} - b) = 
\open \frac{1}{\SLS{P}+\beta} \chi \gamma_5 T^a \frac{1}{\SLS{P} + \beta}
\close
\]
Anti-commuting the $\gamma_5$ through, and using the fact that $\chi$ 
commutes with $p_{\mu}$, gives
\begin{equation}
(a^2 p^2 - b^2) \chi = \open \frac{1}{\SLS{P}+\beta} \chi \frac{1}{\SLS{P}
- \beta} \close
\label{seven}
\end{equation}
If $\chi = \beta$, this equation is exactly satisfied due to the gap equation,
Eq. (\ref{four}), when the quark mass is zero!  The fact that a solution is
possible when $k_{\mu} \rightarrow 0$ means that indeed the pion is 
massless.\footnote{It would appear that the same 
argument should hold for the
U(1) meson, the $\eta'$.  However, for this meson the $1/N$ corrections
connected with the anomaly cannot be ignored \cite{anom}. }
And this approach clearly shows the dual nature of the pion as both a
bound state and as a Goldstone boson:  Eq. (\ref{seven}) is the Bethe-Salpeter
equation for a meson; yet it is identical to the 
gap equation for the vacuum,\footnote{The equivalence between the 
Bethe-Salpeter equation and the gap equation was also shown, for the
case of scalar exchange between fermions, by Migdal and Polyakov
\cite{poly}.} Eq.\ (\ref{four}).  Furthermore, the internal wavefunction of the pion can
be related to a matrix element of the vacuum (for $n_f$ flavors and $N$
colors; $\mbox{tr}'$ is a trace over all indices except spacetime):
\begin{eqnarray*}
\chi(x-y) & = & \beta(x-y) = \frac{-i}{4 n_f N} \mbox{tr}' (G_0) \\
 & = & \frac{-i}{4 n_f N} \mbox{tr}' \langle 0 | \psi(x) \bar{\psi}(y)
 | 0 \rangle .
\end{eqnarray*}

The effective action for pions can now be constructed, using the ansatz of
Eq. (\ref{ansatz}) with $\chi = \beta$, and relaxing the condition 
$\Pi^a(y) = \mbox{const}$.  The quark masses are assumed to be nonzero; 
for simplicity only the two-flavor case, with $m_u = m_d \equiv m$,
will be considered.  Defining $\fatpi = \Pi^a T^a$, 
\begin{eqnarray}
\Gamma & = & \frac{i}{2} \mbox{tr} \left( \beta \gamma_5 \fatpi
 (a \sls{p} - b) \beta \gamma_5 \fatpi (a \sls{p} - b)  \right) \nonumber \\
 & & \, - \frac{i}{2} \mbox{tr} \open \beta \gamma_5 \fatpi 
 \frac{1}{\SLS{P} + \beta} \beta \gamma_5 \fatpi \frac{1}{\SLS{P} + \beta}
 \close + \ldots \; .
\label{gp}
\end{eqnarray}

The action above can be used to determine the mass of the pions in the
presence of explicit chiral symmetry breaking (i.e. when the 
quarks have a small mass).
One can perform
a derivative expansion on the field $\Pi$ \cite{deriv}, commute
the $\gamma_5$ matrices through other quantities until they
stand next to each other, and use the definition of $\beta$ in terms of $a$ and
$b$ to obtain:
\[
\Gamma = - \frac{i}{2} \mbox{tr} (b \beta \fatpi^2) + \frac{i}{2}
 \mbox{tr} \open \frac{1}{\SLS{P}+\beta} \beta \frac{1}{\SLS{P}-\beta} \beta
 \fatpi^2 \close + \frac{1}{2Z} (\partial_{\mu} \Pi^a)^2 + \ldots
\]
where the constant $Z$ can be determined from the
derivative expansion of Eq. (\ref{gp}).  The
gap equation, Eq. (\ref{four}), further simplifies the action:
\[
\Gamma = - \frac{i}{2} \mbox{tr} (m \beta \fatpi^2) + \frac{1}{2Z}
(\partial_{\mu} \Pi^a)^2 + \mbox{(interactions)}
\]
The physical pion field, $\pi_{phys}$, should have the canonical normalization
for its kinetic energy.  Therefore the field $\Pi$ is related to
$\pi_{phys}$ by $\pi_{phys} = \Pi/ \sqrt{Z}$.  Then the mass of the pion
can be read off from the action above (using $\mbox{tr}(T^aT^b) = 
\frac{1}{2} \delta^{ab}$):
\[
m_{\pi}^2 = \frac{i \, m \, \mbox{tr}(\beta) Z}{4} .
\]

The normalization constant $Z$ can be evaluated in the following way:
the Bethe-Salpeter amplitude is $\phi = 
\langle 0 | \psi(x) \bar{\psi}(y) | \pi^a \rangle = 
\beta(x-y) \Pi(y) \gamma_5 T^a$.  Therefore
\begin{eqnarray*}
\langle 0 | \bar{\psi}(0) \gamma_5 T^b \psi(0) | \pi^a \rangle  & = & 
 - \frac{1}{2} \delta^{ab} 4 N \beta(0) \Pi(0) \\
 & = & - \frac{1}{4} \delta^{ab} \mbox{tr}(\beta) \sqrt{Z}
\end{eqnarray*}
On the other hand, current algebra can be used to show that
\[
\langle 0 | \bar{\psi}(0) \gamma_5 T^b \psi(0) | \pi^a \rangle = 
 \frac{i}{2 f_{\pi}} \langle \bar{\psi} \psi \rangle \delta^{ab} =
 \frac{ \mbox{tr}(\beta)}{2 f_{\pi}} \delta^{ab} .
\]
Therefore $ \sqrt{Z} = -2/f_\pi $, and
\[
m_{\pi}^2 =  - \frac{m \langle \bar{\psi} \psi \rangle}{f_{\pi}^2}
\]
The effective action therefore reproduces this classic result for the 
mass of the pion \cite{pokorski}.

\section{Other Mesons}
\label{othermesons}

Encouraged by the effective action's ability to reproduce pion results,
it is time to move on to heavier mesons.  It is
possible, with one additional assumption, to produce another current
algebra result relating the mass of the $\rho$ to the mass of the $a_1$.

The $\rho$ and $a_1$ mesons can be parametrized as
\[
\mbox{\boldmath $\rho$} = \xi(x-y) R^a_{\mu}(y) \gamma^{\mu} T^a \; ; \;\; 
{\bf a}_1 = \zeta(x-y) A^a_{\mu}(y) \gamma^{\mu} \gamma_5 T^a
\]
The extra assumption is that the internal wavefunctions of the $\rho$ and
$a_1$ are nearly equal:  $\xi \approx \zeta$.  The relation between the
$\rho$ and $a_1$ can then be sketched as follows:  using the same method
as for the pions, the part of the effective action that determines the 
mass of the $\rho$ contains a term
\[
- \frac{1}{2 Z_\rho} m_{\rho}^2 \sim
 - \frac{i}{32} \mbox{tr} \open \xi \gamma^\mu \frac{1}{\SLS{P} + \beta}
 \xi \gamma_\mu \frac{1}{\SLS{P} + \beta} \close
\]
where $Z_\rho$ is the normalization constant for the $\rho$-meson.
Similarly, the mass of the $a_1$ contains a piece
\begin{eqnarray*}
- \frac{1}{2Z_a} m_{a_1}^2  & \sim & 
 - \frac{i}{32} \mbox{tr} \open \xi \gamma^\mu \gamma_5 
 \frac{1}{\SLS{P} + \beta}
 \xi \gamma_\mu \gamma_5 \frac{1}{\SLS{P} + \beta} \close \\
 & \sim &  - \frac{i}{32} \mbox{tr} \open \xi \gamma^\mu 
 \frac{1}{\SLS{P} - \beta}
 \xi \gamma_\mu \frac{1}{\SLS{P} + \beta} \close
\end{eqnarray*}
Therefore,
\begin{eqnarray}
\frac{m_{a_1}^2}{Z_a} - \frac{m_{\rho}^2}{Z_\rho} & \sim &
 \frac{i}{16} \mbox{tr} \open \xi \gamma^\mu \left( \frac{1}{\SLS{P} - \beta}
 - \frac{1}{\SLS{P} + \beta} \right) \xi \gamma_\mu \frac{1}{\SLS{P}+\beta}
 \close \nonumber \\
 & = & \frac{i}{8} \mbox{tr} \open \xi \gamma^\mu \frac{1}{\SLS{P}+\beta}
 \beta \frac{1}{\SLS{P}-\beta} \xi \gamma_\mu \frac{1}{\SLS{P}+\beta} 
 \close \nonumber \\
  & = & \frac{i}{8} \mbox{tr} \open \xi \gamma^\mu \frac{1}{\SLS{P}+\beta}
 \beta \gamma_5 \frac{1}{\SLS{P}+\beta} \xi \gamma_\mu \gamma_5
 \frac{1}{\SLS{P}+\beta} 
 \close
\label{threepoint}
\end{eqnarray}
Notice that the last term has the form of a $\rho-\pi-a_1$ interaction
({\it cf}.\  Eq.\ (\ref{mesona})).  Using the complete expressions for the
masses of the $\rho$ and $a_1$ confirms this argument.
Defining the coupling $f_{a \rho \pi}$ by the interaction
\[
{\cal L}_{int} = i f_{a \rho \pi} \epsilon_{abc} \rho^a_\nu a_1^{\nu b}
 \pi^c
\]
we find
\[
\frac{m_{a_1}^2}{Z_a} - \frac{m_{\rho}^2}{Z_\rho} = 
- 2 \frac{f_{a \rho \pi}}{\sqrt{Z_a Z_\rho Z_\pi}}
\]
This sum rule can be put into a more convenient form by using the following
definitions of $g_a$ and $g_\rho$,
\[
\langle 0 | \bar{\psi}(0) \gamma_\nu T^b \psi(0) | \rho^a_\mu \rangle
 = g_\rho \delta^{ab} g_{\mu \nu} = -\frac{1}{4} \delta^{ab} g_{\mu \nu}
 \mbox{tr}(\xi) \sqrt{Z_\rho} 
\]
\[
\langle 0 | \bar{\psi}(0) \gamma_\nu \gamma_5 T^b \psi(0) | a_{1 \mu}^a
\rangle = g_a \delta^{ab} g_{\mu \nu} = -\frac{1}{4} \delta^{ab} g_{\mu \nu}
 \mbox{tr}(\xi) \sqrt{Z_a}
\]
and using $\sqrt{Z_\pi} = - 2/f_\pi$ to obtain
\begin{equation}
\frac{m_{a_1}^2}{g_a^2} = \frac{m_{\rho}^2}{g_{\rho}^2} + \frac{f_\pi
 f_{a \rho \pi} }{g_a g_\rho} 
\label{sumrule}
\end{equation}

This sum rule was originally derived by Schnitzer and Weinberg \cite{sandw}
using current algebra.
In order to test the relation numerically, the masses of the $\rho$ and
$a_1$ can be taken from the Particle Data Group \cite{pdg}; $f_\pi 
\approx 92$ MeV \cite{dgh}; $g_\rho$ can be extracted from $\rho \rightarrow
e^+ e^-$ decay, $g_\rho = 0.117 \pm .005 \, \mbox{GeV}^2$ \cite{dgh};
and $f_{a \rho \pi}$ can be obtained from $a_1 \rightarrow \rho \pi$ decay,
$f_{a \rho \pi} = 4.6 \pm .2$ GeV \cite{isgur}.  
The only problematic quantity is $g_a$.
Assuming that the decay $\tau \rightarrow \rho \pi$ always proceeds through
the $a_1$ resonance, $g_a = 0.177 \pm 0.014 \mbox{GeV}^2$ \cite{isgur}.
However, Weinberg's sum rule \cite{weinberg},
\[
\frac{g_{\rho}^2}{m_{\rho}^2} = \frac{g_a^2}{m_{a_1}^2} + f^2_\pi
\]
gives a value $g_a = 0.15 \mbox{GeV}^2$.  Using the first value of
$g_a$, Eq.(\ref{sumrule}) gives
\[
48 \mbox{GeV}^{-2} = 64 \mbox{GeV}^{-2} \: ,
\]
which is about a 25\% discrepancy.  However, the second value for $g_a$ gives
\[
67.2 \mbox{GeV}^{-2} = 67.4 \mbox{GeV}^{-2} 
\]

\section{Conclusion}

The utility of the effective action is that it provides---as do all 
effective actions---a convenient bookkeeping device for the quantum
numbers of the particles.  However, unlike the purely phenomenological
actions, the terms in this action automatically appear with the correct
symmetry properties, since they are derived formally from QCD.
Furthermore, the bilocal action provides a systematic approach to the study of
mesons, housing the gap equation, the Bethe-Salpeter equation, and
scattering amplitudes all under one roof.  It also provides a 
starting point for more detailed model-based calculations---if one would
like to use a particular ansatz for the Green function $G_0$, for example.
However, as the above results demonstrate, a number of calculations
can be performed even in the absence of such assumptions.

\section*{Acknowledgements}

It is a pleasure to acknowledge many helpful conversations with
Lay Nam Chang, Mahiko Suzuki, and Chia Tze.  This work was supported
by the U.S. Department of Energy Grant No. DE-FG05-95ER40709A.

\setcounter{section}{1}

\section*{Appendix:  Gauge Invariance}

\renewcommand{\theequation}{\Alph{section}.\arabic{equation}}
\setcounter{equation}{0}

This appendix addresses the difficult question of gauge invariance.  It 
turns out that the effective action is explicitly gauge dependent, as is
the Green function $G_0$.  However, it will be shown that meson masses, as
well as physical scattering amplitudes, do not depend on the gauge.  The
proof depends on a generalization of methods used for the scalar effective
potential \cite{nielsen}.  An excellent pedagogical introduction to
the subject is given by Ref.\cite{af}.

We begin with the QCD path integral, with various source terms added.
We first add a bilocal source $K$;
it will be replaced by $G$ in
the Legendre transform.  For convenience, we also need a source $h$
that is independent
of $x$, and sources $Q$ and $R$, which are bilocal.  To condense the
notation, the spacetime
dependence of the variables, when expressed, will be written as indices
rather than as arguments (e.g. $Q_{xy}$ instead of $Q(x,y)$).
The path integral,
including the gauge-fixing term and the ghosts, is
\begin{eqnarray*}
 & & Z[K,h,Q,R] = \int \, D\psi \, D\bar{\psi} \, D\eta \, D\bar{\eta}
 \, DA_\mu \exp i \left( \int dx \, \left\{ \bar{\psi} (i \sls{\partial}
 - \sls{A} - m) \psi - \frac{1}{4 g^2} F^2 \right. \right. \\
 & & \;\;\;\;\;\;\;\; \left. \left. - \frac{1}{2 \alpha} 
 (\partial_\mu A_a^\mu)^2  - \bar{\eta} \partial^\mu D_\mu \eta - \frac{1}{2}
 h \bar{\eta}_a \partial_\mu A^\mu_a \right\} \right. \\
 & & \left. + \int dx \, dy \,
 \left\{ - \bar{\psi}_x ( Q_{xy} i \eta^a_y T^a - i \eta^a_x T^a Q_{xy} )
 \psi_y - \bar{\psi}_x \eta^a_x T^a R_{xy} \eta^b_y T^b \psi_y
 - \bar{\psi}_x K_{xy} \psi_y \right\} \right)
\end{eqnarray*}
The trick is to use the Ward identity that results from the 
fact that the path integral is invariant under
the BRS transformation; this can be converted into 
the statement that physical observables do not depend on the gauge 
parameter $\alpha$.  Under the BRS transformation \cite{pokorski} 
(using the ghost equations of motion to simplify the result),
\[
\delta Z = \int \: D[\mu] \left( \frac{1}{2 \alpha} h (\partial_\mu A^\mu_a)^2
 - \bar{\psi} [K, i\eta^a T^a] \psi - 2 \bar{\psi}(\eta^a T^a Q \eta^b T^b)
 \psi \right) e^{iS} = 0
\]
In this equation the variables are treated as having spacetime
indices, and all indices are summed over.  Now the sources have been
carefully introduced in just such a manner that the above equation can
be rewritten using derivatives with respect to the sources:
\begin{equation}
\left( h \alpha \frac{\partial}{\partial \alpha} + \mbox{tr} (K_{xy}
\frac{\partial}{\partial Q_{xy}} ) + 2 \mbox{tr} (Q_{xy}
\frac{\partial}{\partial R_{xy}} ) \right) Z = 0
\label{dz}
\end{equation}
From here the generator of connected Green functions W is introduced in
the usual way: $Z = \exp(iW)$.  The effective action $\Gamma$ is the
Legendre transform of $W$:
\[
\Gamma[G] = W[K] - \mbox{tr} (GK)
\]
with \[
\frac{\partial W}{\partial K} = G \;\;\;\; ; \;\;\;\; 
 \frac{\partial \Gamma}{\partial G} = -K
\]
The sources $h$, $Q$, and $R$ are not Legendre-transformed.  
Eq.(\ref{dz}) becomes
\begin{equation}
\mbox{tr} \left( - \frac{\partial \Gamma}{\partial G}
 \frac{\partial \Gamma}{\partial Q} + 2 Q \frac{\partial \Gamma}
 {\partial R} \right) + h \alpha \frac{\partial \Gamma}{\partial \alpha}
 = 0
\label{ward1}
\end{equation}
Eq. (\ref{ward1}) is the main Ward identity from which everything else
in this appendix is derived.

To begin, we take the partial derivative of Eq. (\ref{ward1}) with
respect to $h$, set $Q = h = 0$, and omit terms which vanish due to
ghost conservation:
\begin{equation}
- \mbox{tr} \left( \frac{\der \Gamma}{\der G} \frac{\der^2 \Gamma}
 {\der h \der Q} \right) + \alpha \frac{\der \Gamma}{\der \alpha}
 = 0
\label{ward2}
\end{equation}
This equation states that the action changes when $\alpha$ is
changed. However, the value of $\Gamma$ at the point where
$\der \Gamma/\der G = 0$ does not in fact depend on $\alpha$.

To learn about the meson two-point function, we take the derivative
$\der^2/\der G_{xy} \der G_{zw}$ of Eq. (\ref{ward2}).  (When
$G$ and $Q$ appear below without (spacetime) indices, the indices of
$G$ are contracted with the indices of $Q$.)
\begin{eqnarray}
 & & \left(- \frac{\der^2 \Gamma}{\der h \der Q} \frac{\der}{\der G}
 + \alpha \frac{\der}{\der \alpha} \right) \frac{\der^2 \Gamma}
 {\der G_{xy} \der G_{zw}} - \frac{\der^3 \Gamma}{\der h \der Q \der G_{xy}}
 \frac{\der^2 \Gamma}{\der G \der G_{zw}} \nonumber \\
 & & - \frac{\der^2 \Gamma}{\der G_{xy} \der G} \frac{\der^3 \Gamma}
 {\der h \der Q \der G_{zw}} - \frac{\der \Gamma}{\der G} 
 \frac{\der^4 \Gamma}{\der h \der Q \der G_{xy} \der G_{zw} }
 = 0
\label{ward3}
\end{eqnarray}
Two observations simplify this equation somewhat.  
First, this equation should be evaluated
at $G=G_0$, where $\der \Gamma/\der G = 0$, so
the last term of Eq. (\ref{ward3}) vanishes.
Second, it can be shown\footnote{The analogous case holds for the
scalar potential \cite{nielsen,af}.  In this context it can be proved
by taking the derivative of Eq. (\ref{dz}) with respect to $K$,
Legendre transforming the result, and setting $G=G_0$.} that the function 
$\der^2 \Gamma/\der h \der Q$, evaluated at $G=G_0$,
is equal to $- \alpha \, \der G_0/\der \alpha$.
Therefore the expression in the first term of Eq. (\ref{ward3}),
\[
\left. \alpha \frac{\der}{\der \alpha} - \frac{\der^2 \Gamma}{\der h \der Q}
 \frac{\der}{\der G} \right|_{G=G_0} =
 \alpha \left( \frac{\der}{\der \alpha} + \frac{\der G_0}{\der \alpha}
 \frac{\der}{\der G_0} \right) \equiv \alpha \frac{D}{D \alpha}
\]
gives the total $\alpha$-dependence, both implicit and explicit.

Eq. (\ref{ward3}) shows how the inverse two-point function changes
as $\alpha$ changes; we also need to know how the Bethe-Salpeter
wavefunction changes.
We can multiply Eq. (\ref{ward3}) by the Bethe-Salpeter
wavefunction $\phi_{zw}$ from the right, and use the Bethe-Salpeter
equation $( \der^2 \Gamma/\der G_{uv} \der G_{zw}) \phi_{zw} = 0$
(sum over repeated indices!) to get
\[ 
\left( \alpha \frac{D}{D \alpha} \frac{\der^2 \Gamma}{\der G_{xy} \der G_{zw} } 
\right) \phi_{zw} - \frac{\der^2 \Gamma}{\der G_{xy} \der G}
\frac{\der^3 \Gamma}{\der h \der Q \der G_{zw}} \phi_{zw} = 0
\]
Under a change of gauge, $\alpha \rightarrow \alpha + \delta \alpha$,
the wavefunction $\phi$ must change by a certain amount 
$\delta \phi = (D\phi/D \alpha) \delta \alpha$
in order for it to still satisfy the Bethe-Salpeter equation.  The
amount by which it must change can be read off from the above equation:
\begin{equation}
\alpha \frac{D}{D \alpha} \phi_{zw} =
 - \frac{\der^3 \Gamma}{\der h \der Q_{zw} \der G_{z'w'}} \phi_{z'w'}
\label{dphi}
\end{equation}
This ensures that
\[
\alpha \frac{D}{D \alpha} \left[ \frac{\der^2 \Gamma}{\der G_{xy}
 \der G_{zw}} \phi_{zw} \right] = 0
\]
which means that indeed $\phi + \delta \phi$ is the correct
Bethe-Salpeter wavefunction in the new gauge.

It proves more convenient to express Eq. (\ref{dphi}) in momentum space.
Recall that the Bethe-Salpeter wavefunction can be factored:
$\phi_{xy} = \chi(x-y) H(y)$ where $\chi$ describes the internal
motion, and $H$ the center-of-mass motion.  Fourier transforming
this relation gives $\phi(q;P) = \chi(q) H(P)$, where $P$ obeys
the relation $P^2 = M^2$.  Now writing the
Fourier transform of the function appearing in Eq.(\ref{dphi}) as
\[
\frac{\der^3 \Gamma}{\der h \der Q_{xy} \der G_{zw}}
= \int dq \, dk \, dP' \,\tilde{\Gamma}^{(3)}(q,k;P') e^{iq.(x-y)} e^{ik.(z-w)}
 e^{iP'.(y-w)}
\]
Eq. (\ref{dphi}) becomes
\[
\alpha \frac{D}{D \alpha} \left( \chi(q) H(P) \right)
 = - \int dk \, \tilde{\Gamma}^{(3)}(q,k;P) \chi(k) H(P)
\]
(Here translation invariance has forced $P'$ to equal $P$.)
The function $H$ does not change under the gauge transformation,
and we have
\begin{equation}
\alpha \frac{D}{D \alpha} \chi(q) = - \int dk \, 
 \left. \tilde{\Gamma}^{(3)} (q,k;P) \right |_{P^2 = M^2} \chi(k)
\label{dchi}
\end{equation}

At last we are ready to show that meson masses are gauge independent.
$H$ is a local
function, and we can regard $\Gamma$ as a functional of $H$,
so that the derivatives of $\Gamma$ with respect to $H$ give the
one-particle irreducible amplitudes for meson scattering. To do this,
we write $H = H(P)$, where $P$ can now take on any value.
Then the Fourier transform of the inverse two-point function is
\[
\frac{\der^2 \Gamma}{\der H^2} (P) = \int dq \, dp \, 
 \chi(p) \frac{\der^2 \Gamma}{\der G^2} (p,q;P) \chi(q)
\]
Using Eq. (\ref{ward3}) and Eq. (\ref{dchi}) we find that
\[
\alpha \frac{D}{D \alpha} \left(  \left. \frac{\der^2 \Gamma}{\der H^2}
 (P) \right|_{P^2 = M^2} \right) = 0
\]
That is, if $P^2 = M^2$, the value of the inverse two-point function 
does not depend on the gauge parameter $\alpha$. Therefore the
condition $P^2 = M^2$, which causes the inverse two-point function
to vanish, causes it to vanish in all gauges.  This means that
the value of the (renormalized) physical mass $M$, defined as the
point where the inverse two-point function vanishes, does not depend
on the choice of gauge.

The three-point functions can be analyzed in a similar way.  We take
the derivative $\der/\der G_{uv}$ of Eq. (\ref{ward3}); multiply by
the wavefunction $\chi(k)$ where $k$ is the momentum conjugate to
$(u-v)$; then use Eq. (\ref{dchi}) to find
\[
\alpha \frac{D}{D \alpha} \left( \frac{\der^3 \Gamma}
 {\der H^3} (P,Q; -P-Q) \right) = 0
\]
if all three particles are on shell.  That is, the off-shell amplitude
in general depends on the gauge, but the physical three-point
scattering amplitude is gauge independent.
 
The three-point function calculated in section \ref{othermesons} 
was not
taken on shell, but with the external momenta going to zero.  However,
Eq. (\ref{threepoint}) is gauge {\it co-}variant, by construction.
It relates the masses to the wavefunction renormalization constants
and the three-point function, all evaluated at $p=0$.  The subsequent
derivation of Eq.(\ref{sumrule}) relies on the additional assumption,
common in meson physics, that the $Z$'s  and $f_{a \rho \pi}$
depend only weakly on the momentum, no matter what the gauge.

Finally, there is an amusing property of the four-point function, 
a Ward identity for which
can be obtained by taking the derivative $\der^2/\der G_{uv} \der G_{st}$
of Eq. (\ref{ward3}).  Curiously, the one-particle
irreducible four-point function is {\it not} gauge-invariant, even
when the external particles are on-shell.  Only the complete
four-point amplitude---the sum of irreducible and
reducible diagrams---is gauge invariant.  The fact that the irreducible
four-point function mixes with the reducible functions under a
gauge transformation means that a meson cannot be thought of as a
distinct particle that maintains its identity throughout the reaction!

\section*{Figure Captions}

\vskip 1cm

%\verb| |

{\bf Fig. 1} The two-particle irreducible diagrams contributing to the
effective action.

\end{document}